\documentclass[preprint,preprintnumbers,amsmath,amssymb]{revtex4}

\usepackage{comment}
\usepackage{amssymb}
\usepackage{amsmath}
\usepackage[dvipsnames,usenames]{color}
\usepackage{graphicx,graphics}
\usepackage{amssymb}

\begin{document}
\title{On-demand Entanglement Source with Spatial Phase Modulation}
\author{Xiang-Bin Wang}
 \email{xbwang@mail.tsinghua.edu.cn}
\author{Cheng-Xi Yang}
\affiliation{Department of Physics and  the Key Laboratory of Atomic
and Nanoscien ces, Ministry of Education, Tsinghua University,
Beijing 100084, China}
\author{Yan-Bing Liu}
\affiliation{Department of Physics and  the Key Laboratory of Atomic
and Nanoscien ces, Ministry of Education, Tsinghua University,
Beijing 100084, China}

\date{\today}

\begin{abstract}
  The polarization entanglement photon pairs
   generated from the biexciton cascade decay in a
  single semiconductor quantum dot is corrupted by the position-dependent (time-dependent) phase difference
  of the two polarization mode due to the fine structure splitting. We show that,
  by taking voltage ramping to an electro-optic modulator, such
  phase-difference can be removed.
  In our first proposed set-up, two photons
  are sent to two separate Pockels cell under reverse voltage ramping, as a result, the position-dependent
  phase difference between the two polarization mode is removed in the outcome
  state. In our second proposed set-up, the polarization of the first photon
  is flipped and then both photons fly into the same Pockels cell.
  Since we only need to separate the two photons rather than
  separate the two polarization modes, our schemes are robust
  with respect to fluctuations of the optical paths.

\end{abstract}

\maketitle \pagenumbering{arabic}

\section{ Introduction} Quantum entanglement plays an important role
in the study of fundamental principles of quantum
mechanics\cite{sakurai}. It is also the most important resource in
quantum information processing\cite{nielsen,decoy}. Among all types
of quantum entanglement, polarization entangled photon-pairs are
particularly useful because of easy manipulation and transmission.
There are many mature techniques to produce such entangled pairs
\emph{probabilistically}\cite{PRL_93_KiessTE,
  PhysRevLett.75.4337, Nature_04_EdamatsuK, PRL_04_FattalD},
while an \emph{on-demand} entangled photon pair is essential in many
tasks in quantum information processing.

Recently, an on-demand entangled photon-pair source was
proposed\cite{PRL_00_OliverB} and realized in a semiconductor
quantum dot system\cite{Nature_06_StevensonRM, AkopianN_PRL_06,
NatPhon_ShieldsAJ_07}. However, because of the fine-structure
splitting (FSS) there, the relative phase of the entangled state is
randomized so that only classical correlation can be detected by
traditional time-integrated
measurement\cite{stace2003,PRL_07_HudsonAJ,PRL_08_StevensonRM,guo}.
So far, there are many methods proposed to explore this ``hidden
entanglement''\cite{stace2003,PRL_07_HudsonAJ,PRL_08_StevensonRM,guo,
NJP_06_YoungRJ,NJP_07_HafenbrakR,he:157405,
  PRL_09_YoungRJ,jones},
for example, reducing
FSS\cite{NJP_06_YoungRJ,NJP_07_HafenbrakR,he:157405,
  PRL_09_YoungRJ}, spectral filtering\cite{AkopianN_PRL_06}, time
resolving post-selection\cite{PRL_08_StevensonRM}, and so on. Up to
now, the smallest FSS realized in experiment is about $0.3\ \mu eV$
and non-classical nature of the radiation field is verified by
directly observing violation of the Bell
inequality\cite{PRL_09_YoungRJ}. However, the entanglement quality
is considerably decreased even by very small FSS, and further
reducing FSS is very difficult in experiment. Furthermore, the
severe restriction on FSS greatly limits the selection range of
quantum dot systems. Certain quantum dots with large FSS cannot be
used even if they have distinct advantage, such as emitting photons
of frequencies in the easy transmission frequency window in free
space or optical fiber. Also, the post-selection method in frequency
domain or time domain will significantly decrease the photon
collection efficiency. To overcome all these drawbacks, Stace et al
proposed to use cavities to control the frequencies. Latter, Jones
and Stace proposed a more simplified, downstream solution of
polarization-dependent frequency shift to the photons by an
acousto-optical modulator (AOM). Basically, there are 3 steps in the
circuit: 1) Split the polarization modes of each photons; 2) Shift
the frequency of vertical polarization modes by AOM; 3)Combine split
beams by a polarization beam splitter. However, the efficiency
of a normal commercially available AOM is fairly low. The efficiency
of a very good AOM for a single photon is about $80\%$. The joint
efficiency of two photons in the scheme in Ref.\cite{jones} is not
larger than $64\%$. Moreover, in the proposed set-up\cite{jones},
special care has to be taken to make the two optical paths  stable.
Say, a fluctuation of half a micro meter in any optical path will
entirely destroy the result.

 Here we propose another solution through using an electro-optical modulator (EOM).
EOM is a mature technique which has been demonstrated in many
experiments. In particular, two-photon interference has been
experimentally observed\cite{np} very recently.
 In our proposed set-up, we use a dichroic mirror to separate the two photons, and
then remove the position-dependent phase difference by using
Pockels cells under a ramping voltage.
 Since a Pockels cell itself makes the phase shift differently to
 different polarization mode, we don't have to separate the
 different polarization modes as was proposed in Ref.\cite{jones}.
 Instead, we only need to separate the two photons. In this way, compared with the existing
proposal\cite{jones}, our method has an advantage in its robustness
to fluctuations of optical paths. As calculated later, a
fluctuation of 1 mm will cause $10^{-3}$ fluctuation in the phase.
Moreover, compared with Ref.\cite{jones}, our scheme seem to have a significantly higher efficiency.
A commercially available Pockels cell almost has no loss.

\section{ The problem}
\begin{figure}
\includegraphics{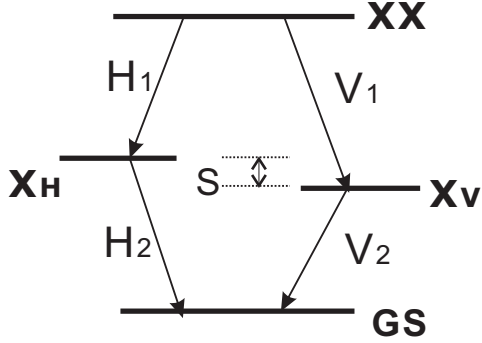}
\caption{\label{fig:levels} Energy levels of the
  semiconductor quantum dot used to generate polarization
  entangled photons. The biexciton state (XX) is a
  zero-spin state formed by two electrons and two heavy
  holes. When the dot decays, two photons are emitted
  sequentially, and their polarization is determined by the
  ``decay path''. Usually an FSS $S$ exists between the two
  excitons ($X_H$) and ($X_V$).}
\end{figure}

 The energy levels of the quantum dot used for photon-pair
generation are shown in Fig.~\ref{fig:levels}. After exciting a
single quantum dot into biexciton state (XX), two photons are
emitted sequentially as the dot decays in a cascade process.
Because the two exciton states ($X_H$ and $X_V$) are not
degenerate\cite{PhysRevLett.76.3005,PhysRevB.65.195315}, the two
photons are actually entangled in the complex space of both
polarization and frequency
\begin{equation}
  \label{eq:stateFre}
  \begin{split}
  |\Psi\rangle =& \frac{1}{\sqrt{2}}
   \bigg[ \iint_{-\infty}^{\infty} d\omega_1 d\omega_2
    \Phi_H(\omega_1,\omega_2) |H_1H_2;\omega_1,\omega_2\rangle\\
   & + \iint_{-\infty}^{\infty} d\omega_1 d\omega_2
    \Phi_V(\omega_1,\omega_2) |V_1V_2;\omega_1,\omega_2\rangle
  \bigg].
  \end{split}
\end{equation}
The spectral functions for the two decay path of the quantum dot
system can be written as\cite{AkopianN_PRL_06, scully}
\begin{subequations}
  \label{eq:13}
  \begin{align}
    \Phi_{H}(\omega_1,\omega_2) =& \frac{\sqrt{2}\Gamma}{2\pi}
  \frac{1}{\omega_1 + \omega_2 - \omega_{0} + i\Gamma}\nonumber\\
  & \times\frac{1}{\omega_2 - \omega_{H_2} +
    i\Gamma/2},\label{eq:14} \\
  \Phi_{V}(\omega_1,\omega_2) =& \frac{\sqrt{2}\Gamma}{2\pi}
  \frac{1}{\omega_1 + \omega_2 - \omega_{0} + i\Gamma}\nonumber\\
  & \times\frac{1}{\omega_2 - \omega_{V_2} +
    i\Gamma/2}.\label{eq:20}
  \end{align}
\end{subequations}
Here, as shown in Fig.~\ref{fig:levels},
$\omega_{H_2}=\omega_{X_H}-\omega_{GS}$, $\omega_{V_2}=\omega_{X_V}
-\omega_{GS}$, and $\omega_0 = \omega_{XX}-\omega_{GS}$, where
$\hbar\omega_{XX}$, $\hbar\omega_{X_H}$, $\hbar\omega_{X_V}$,
$\hbar\omega_{GS}$ are the eigenenergy of levels $XX$, $X_H$, $X_V$,
and $GS$, respectively, and $\Gamma$ is the decay rate of the four
transitions $XX\rightarrow X_H$, $XX\rightarrow X_V$,
$X_H\rightarrow GS$, and $X_V\rightarrow GS$\cite{AkopianN_PRL_06}.

Therefore, the state is actually inseparable in the composite space
of both polarization and frequency. This hides the entanglement in
polarization space only.

\section{ Phase modulation with Pockels cells} 
\begin{figure}
  \includegraphics[width=8cm]{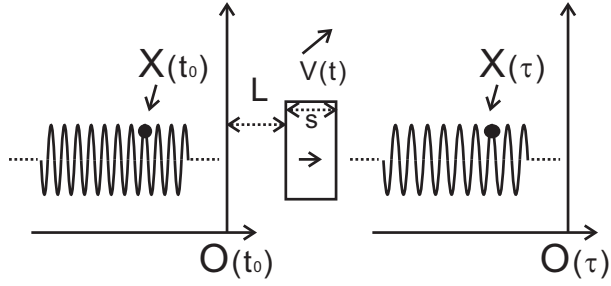}
  \caption{\label{fig:modu} Phase modulation. After the wave
    train passed the crystal (the square box in the middle),
    the point $X(t_0)$ propagated to $X(\tau)$. The relationship of
    the positions of $X$ and $X'$ is given in Eq.~(\ref{xtau}).
  }
\end{figure}

A Pockels cell contains a
crystal whose refraction index of a certain optical axis changes
linearly with the external voltage, due to the so called
electro-optical effects. Consider a Pockels cell with a
time-dependent voltage $V(t)$ and a wave packet passing through it,
as shown in Fig.~\ref{fig:modu}. For simplicity, we assume that any non-trivial
phase modulation only happens to the vertical polarization of the
incident light.

Suppose initially the wave function of a wave train in vertical
polarization is $e^{ik_Vx}$, with a certain reference original point
$O(0)$, at the left side of the crystal. We shall always use
the reference framework of the flying wave train itself, i.e., the
reference point $O(t)$ propagates with the wave train, in the same
speed.  Suppose at time $t_0$, the distance between the reference
point $O(t_0)$ and the left side surface of the crystal is $L$. The
ramping voltage $V(t)$ applied to the crystal is a linear function
of time $t$, $V(t)=a+bt$. Suppose the position of any point $X(t_0)$
is $x(t_0)$ at this reference framework. At a later time $t=\tau$,
the wave train is at the right side of the crystal and the phase of
original points $O(t_0),\;X(t_0)$   have now propagated to points
$O(\tau),\;X(\tau)$, respectively. At time $\tau$, we take $O(\tau)$
as the reference original point and denote $x(\tau)$ as the position
of $X(\tau)$ in the new reference framework of $O(\tau)$. To see the
phase modulation after the wave train passes the crystal, we study
the relation between $x(t_0)$ and $x(\tau)$.

The refraction index of the crystal is linearly dependent on the
applied voltage. At any time $t$, the vertical-polarization-mode
light speed inside the crystal is
\begin{equation}
v(t)=\frac{v_0}{1+\eta V(t)}
\end{equation}
and $v_0$ is the light speed inside the crystal when there is no
applied voltage, $\eta$ is a constant parameter which is dependent
on the crystal property itself.
Suppose the
crystal thickness is $s$.   At time $t_0$, the original phase at
point $O(t_0)$ is $\varphi_{OV}$ or $\varphi_{OH}$, for vertical
polarization wave train or horizontal polarization wave train,
respectively.

{\em Frequency shift.} Consider the vertical polarization case
first. Suppose it takes time $\Delta t(X)$ for point $X$ to pass
through the crystal. Explicitly,
\begin{equation}\label{dt}
\int_{t_{in}(X)}^{t_{in}(X)+\Delta t(X)} v(t)dt = s
\end{equation}
where $t_{in}(X)=\frac{-x}{c}+L/c$ is the time point that point $X$
in the original wave train reaches the left side of the crystal.
For a linearly rising voltage $V(t)=a+bt$, Eq.(\ref{dt})
gives rise to
\begin{equation}\label{time}
\Delta t (X) = \frac{1+\eta(a+bL/c - b x/c)}{\eta b} \left(e^{\eta
bs/v_0}-1\right).
\end{equation}

At time point $\tau$, the phases of points $O(t_0),\;X(t_0)$ have
propagated to points $O(\tau),X(\tau)$, respectively. Using the
formula above we find that the position of $X(\tau)$ in the new
reference framework $O(\tau)$ is
\begin{equation}\label{xtau}
x(\tau)=e^{\eta bs/v_0}x.
\end{equation}
This is to say, after passing through the crystal, the spatial phase
function (with reference original point $O(\tau)$) is changed into
\begin{equation}
\varphi_{out} (x(\tau))=\varphi_{out}(e^{\eta bs/v_0}x(t_0)) =
\varphi_{in}(x(t_0))
\end{equation}
where $\varphi_{in}(x)=k_Vx$ is the spatial phase function of the
wave train before passing through the crystal, with reference
original point $O(t_0)$. Therefore, the spatial phase  for the wave
train after passing through the crystal is
\begin{equation}
\varphi_{out} (x) = e^{-\eta bs/v_0}kx,
\end{equation}
where the reference original point is $O(\tau)$. (For simplicity,
here we set all initial phases at reference point to be 0.)
 This is actually a
frequency shift to the wave train.
 The crystal under voltage ramping of
$V(t)=a+ bt$ will transform the original frequency $\omega_V$ (or
wave vector $k_V$ ) of the wave train at the left side of the
crystal into the new frequency $\omega_V'$ (or wavevector $k_V'$ )
of the wave train at the right side of the crystal by the following
formula:
\begin{equation}\label{vip}
\frac{\omega_V'}{\omega_V}=\frac{k_V'}{k_V}=e^{-\eta bs/v_0}=f(b).
\end{equation}

{\em Phase change.} Eq.(\ref{vip}) is the spatial phase modulation
of vertical polarization only.
 If the original wave train is in horizontal
polarization mode, it takes time
\begin{equation}
\Delta t_H  =s/v_0
\end{equation}
for any point $X(t_0)$ in the original wave train to pass through
the crystal. At time $\tau$, the original reference point $O(t_0)$
propagates to the new reference point $O_H(\tau)$. In this new
reference point, the spatial phase function is
\begin{equation}
\varphi_H (x_H(\tau)) =k_H x_H(\tau),
\end{equation}
where $k_H$ is the wave vector of the horizontal polarized mode.
At the reference framework of $O(\tau)$ (reference point of vertical
polarization), the position of $O_H(\tau)$ is
\begin{equation}\label{dis}\begin{split}
d(a,b,L)  &= c(\tau -\frac{s}{v_0}) - c(\tau - \Delta
t(O))\\&=\left(\frac{c}{\eta b}+\frac{ac}{b}+L\right)\left(e^{\eta
bs/v_0}-1\right)-\frac{cs}{v_0}\end{split}
\end{equation}
where $\Delta t(O)$ is given by Eq.(\ref{time}) with $x=0$.
Therefore, in the same reference framework $O(\tau)$, the spatial
phase of horizontal polarization at time $\tau$ is
\begin{equation}\label{ph}
\varphi_H (x) =k_H (x-d(a,b))
\end{equation}
where $k_H$ is the wave vector of the horizontal polarization mode
and $d(a,b) $ is given by Eq.(\ref{dis}). The spatial phase
difference of two polarizations in reference framework $O(\tau)$ is
\begin{equation}\label{phout}
\Delta \varphi(x) = \left(e^{-\eta bs/v_0}k_V -k_H\right) x + k_H
d(a,b).
\end{equation}
 From this we can see that the crystal under a time dependent
 voltage not only changes the frequency, but also offers a
 position-independent
  phase difference between the two polarization modes of the outcome wave train.
This phase is dependent on the parameters $a,\,b$ in the linear
function $V(t)$.

In the derivation above, we have ignored the possible small
dispersion of the crystal. Obviously, our result can be directly
extended to this case that the crystal's refraction index is
dependent on the frequency of the incident light by setting $v_0$
and $\eta$ frequency dependent.

For simplicity, we shall only consider the case without dispersion
hereafter. In such a case,  taking a very similar derivation, we
have the following wavefunction transform formulas for arbitrary
wavefunction by its polarization mode:
\begin{equation}\label{tr}\begin{split}
&\psi_V(x) \longrightarrow e^{-\eta bs/(2 v_0)}\psi_V (e^{-\eta bs/v_0}x);
\\
&\psi_H(x)\longrightarrow \psi_H(x-d(a,b)).
\end{split}
\end{equation}

\section{Our scheme}
We propose two schemes here.

\begin{figure}
  \includegraphics{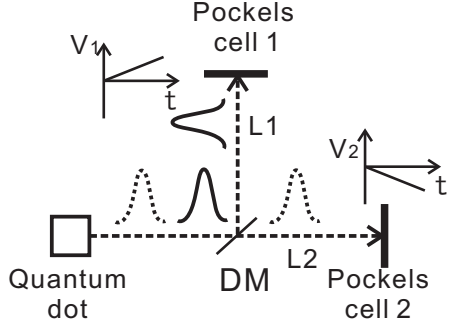}
  \caption{\label{fig:setup} Proposed scheme 1.
  The first photon and second photon are separated by a
    dichroic mirror (DM). The two Pockels cells start to run before
    the photons arrive. They make reverse phase modulation.
  }
\end{figure}

\begin{figure}
  \includegraphics[width=7cm]{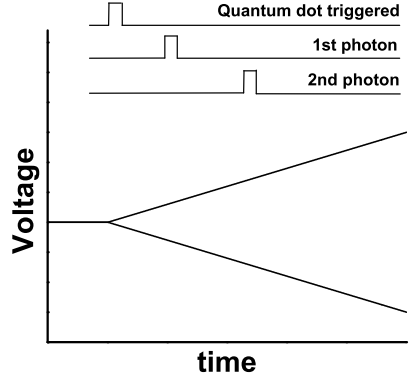}
  \caption{\label{fig:Torder} Voltage ramping in scheme 1.
  }
\end{figure}

As shown in Fig.\ref{fig:setup}, scheme 1 contains two EOM phase
modulators with ramping voltage of $V_1(t)=a_1+b_1t$ and
$V_2(t)=a_2+b_2t$. The two photons are separated by their frequency,
and then pass through each modulator. Each voltage ramping covers
the time that each photon pass through its modulator, as shown in
Fig.~\ref{fig:Torder}. Scheme 1 has no limit to the frequencies of the two
photons, say, no matter they are quite close or quite different.
However, as shown below, we need the starting time difference of two
ramping voltages be controlled much smaller than 1 ns.

\begin{figure}
  \includegraphics[width=6cm]{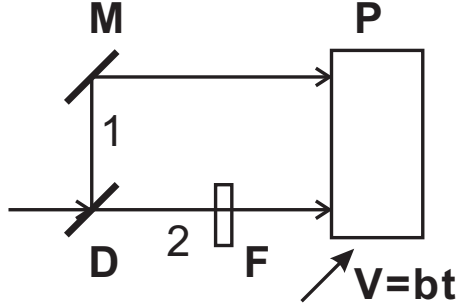}
  \caption{\label{sm} Proposed scheme  2.
   The first photon and second photon are separated by a
    dichroic mirror (D). Polarization of photon 1 is flipped
    by a flipper (F) before enters the Pockels cell P.
    Here both photons enter the same Pockels cell.
  }
\end{figure}

If the frequency difference of two photons is rather small compared
with the frequencies of each photons, we can use Scheme 2. In scheme
2, the problem of starting time difference control is circumvented.
Scheme 2 contains only one EOM phase modulator under voltage ramping
of $V(t)=a+bt$, as shown in Fig.(\ref{sm}). Two photons are
separated and they pass through the modulator in different path.
Polarization of photon one is flipped before it reaches the
modulator. Also, the voltage ramping covers the time of both photon
passing through the modulator.

\subsection{ Voltage ramping of scheme 1} There are two photons
emitted from the quantum dot. If we transform
Eq.~(\ref{eq:stateFre}) from frequency space to position space, the
state of the field can be rewritten as
\begin{equation}
  \label{eq:evo}
  \begin{split}
  |\Psi_{in}\rangle =& \frac{\Gamma}{c}
  \iint_{0>x_1>x_2} dx_1 dx_2
   e^{\frac{\Gamma}{2c}(x_2+x_1)} \\
  & \times\big(e^{i(k_{H_1} x_1 +
    k_{H_2} x_2)} |H_1H_2\rangle
  + e^{i(k_{V_1} x_1 +
    k_{V_2} x_2)} |V_1V_2\rangle\big),
  \end{split}
\end{equation}
where $x_1$ and $x_2$ refer to the position of the first photon and
second photon, respectively, with a common reference point, the
right end of the wave packet. Eq.~\eqref{eq:evo} is equivalent to
the result given in Ref.~\cite{PRL_07_HudsonAJ}. Suppose at a
certain time $t_0$, the reference point $O(t_0)$ arrives at  the
dichroic mirror. According to Eq.(\ref{tr}), the outcome state is
\begin{equation}
  \begin{split}
  |\Psi_{out}\rangle =& \frac{\Gamma}{c} \iint_{0>x_1-d_1>x_2-d_2}
  dx_1 dx_2  A_{H}|H_1H_2\rangle\\
  & \\
    & + \frac{\Gamma}{c}\sqrt{f_1f_2} \iint_{0>f_1x_1>f_2x_2}
  dx_1 dx_2  \\
  &\times A_V
    e^{i\Delta\varphi_1+i\Delta\varphi_2}
    |V_1V_2\rangle,
  \end{split}
\end{equation}
where $f_i=e^{-\eta b_is/v_0}$, $\Delta\varphi_i$ is given by
Eq.(\ref{phout}), with parameters $a=a_i,\;b=b_i,\;L=L_i$ there;
\begin{equation}\begin{split}&
A_H=e^{\frac{\Gamma}{2c}(x_1 + x_2 - d_1 - d_2)}\\& A_V=e^{\frac{\Gamma}{2c}(
f_2x_2+f_1x_1)}
\end{split}
\end{equation}
and $d_i$ is given by Eq.(\ref{dis}) with parameters $a=a_i,b=b_i$.
If we set
\begin{equation}\label{vol}
  \begin{split}
b_1=\frac{v_0}{\eta s}\ln \left( \frac{k_{V_1}}{k_{H_1}} \right)\\
b_2=\frac{v_0}{\eta s}\ln \left( \frac{k_{V_2}}{k_{H_2}}\right),
  \end{split}
\end{equation}
we find that
\begin{equation}\label{phase}\begin{split}
\Delta \varphi_{1} +\Delta\varphi_2& =  k_{H_1}\left(\frac{c}{\eta
b_1}+\frac{ca_1}{b_1}+L_1\right)\left(e^{\eta b_1s/v_0}-1\right)\\
&+ k_{H_2}\left(\frac{c}{\eta
b_2}+\frac{ca_2}{b_2}+L_2\right)\left(e^{\eta b_2s/v_0}-1\right)
-k_0cs/v_0.
\end{split}
\end{equation}
where $L_1,\; L_2$ are the optical path from $O(t_0)$ to the two
separate Pockels cells, $k_0 = k_{H_1} + k_{H_2} $ is a constant
(i.e., position independent). Therefore the value above is a
constant phase independent of positions $x_1,\;x_2$. Also, we find
that
\begin{equation}
\sqrt{f_1f_2}A_V/A_H=\sqrt{f_1f_2}\exp{\left[\frac{\Gamma}{2c}(
f_2x_2+f_1x_1)-\frac{\Gamma}{2c}(x_1+x_2-d_1
-d_2)\right]}=1+\epsilon
\end{equation}
 where $\epsilon$ is in the magnitude order of $10^{-6}$, given the coherence
 length of the wave train
 $L\approx 0.3$m and $\Gamma\approx 10^{9}$.
 Therefore, with the setting of Eq.(\ref{vol}), our scheme 1 can produce high
 quality polarization entangled photon pairs.

In our schemes, we only split the two photons by frequency
difference instead of splitting the polarization mode. Even though
the optical paths of each photons may fluctuate significantly, the
result only changes negligibly. This is different from the AOM based
scheme in Ref.\cite{jones} which separates two polarization
modes.

A fluctuation of amount $\delta l_i$ in the optical path of the
$i$'th photon will cause a fluctuation of amount
\begin{equation}
\delta\varphi_i = k_{H_i}\delta l_i(e^{\eta b_i s/v_0}-1)
\end{equation}
This means, even a fluctuation of 1 mm in one of the optical path
will cause only a phase difference fluctuation of in the magnitude
order  of $10^{-3}$ in our scheme.

Since $a_1,a_2$ plays no role in our scheme, we can ramp the voltage
from 0, i.e., setting $a_1=a_2=0$.  We can also consider the
consequence of non-exact simultaneous voltage ramping of the two
modulators. Suppose the starting times of ramping are $t_1$ and
$t_2$, respectively. To be sure that the voltage ramping covers the
incident wave train, we need $t_1\le t_0$ and $t_2\le t_0$. This is
equivalent to set $a_1=b_1(t_0-t_1)$ and $a_2=b_2(t_0-t_2)$ and
start the voltage ramping exactly at $t=t_0$. The fluctuation in the
value of Eq.(\ref{phase})is now
\begin{equation}\begin{split}
\delta(\varphi) &\approx  \frac{c\eta s}{v_0}[k_{H_1} b_1 (t_0-
t_1)+k_{H_2}b_2(t_0-t_2)] \approx c  k_S \delta t
\end{split}
\end{equation}
where $ k_S =k_{V_1}-k_{H_1}$ which correspond to  FSS, and $\delta
t = t_2-t_1$. We see that out result only dependent on the time
difference $\delta t$, independent of the absolute time. Therefore,
the trigger time uncertainty of the quantum dot does not affect the
result here. To obtain high quality entanglement, we need
\begin{equation}
\delta t << \frac{1}{ck_S}.
\end{equation}
Given an FSS of 1 GHz, we need the ramping time difference to be
much smaller than 1 ns. The consequence of time difference $\delta
t$ here is equivalent to the time window post-selection scheme with
time resolving detection\cite{PRL_08_StevensonRM}. However, here in
our scheme there is almost no photon loss.

In scheme 1,  we need to control the  time difference of two voltage
ramping in a rather small range (much less than 1 ns).  As shown
below, our scheme 2 has an intrinsic fault tolerance property, where
the technical problem of simultaneous ramping is bypassed.

\subsection{Robustness of scheme 2}
In our scheme 2 as shown in Fig.(\ref{sm}), we can set $V(t)=bt$ and
\begin{equation}
  \label{b}
  b=\frac{v_0}{\eta s}\ln(\frac{k_{H_1}}{k_{V_1}}),
\end{equation}
Before the photons pass through each crystal, the state is
\begin{equation}
  \begin{split}
  |\Psi_{in}\rangle =& \frac{\Gamma}{c}
  \iint_{0>x_1>x_2} dx_1 dx_2
   e^{\frac{\Gamma}{2c}(x_2+x_1)} \\
  & \times\big(e^{i(k_{H_1} x_1 +
    k_{H_2} x_2)} |V_1H_2\rangle
  + e^{i(k_{V_1} x_1 +
    k_{V_2} x_2)} |H_1V_2\rangle\big),
  \end{split}
\end{equation}
After the two photons pass through the same Pockels cell under
voltage ramping, the state is
\begin{equation}
  \label{single}
  \begin{split}
  |\Psi_{out}\rangle =& \frac{\Gamma}{c} \sqrt{f}\iint_{0>f x_1>x_2-d_2}
  dx_1 dx_2  A_{H}|V_1H_2\rangle\\
  & \\
    & + \frac{\Gamma}{c} \sqrt{f} \iint_{0>x_1-d_1>f x_2}
  dx_1 dx_2  \\
  &\times A_V
    e^{-i\Delta\varphi_1+i\Delta\varphi_2}
    |H_1V_2\rangle,
  \end{split}
\end{equation}
where
\begin{equation}\begin{split}&
\Delta \varphi_1 = \left(f k_{H_1} - k_{V_1}\right) x_1 +
k_{V_1}d_1,\\
& \Delta \varphi_2 = \left(f k_{V_2} - k_{H_2}\right) x_2 +
k_{H_2}d_2,\\
\end{split}
\end{equation}
with $f = e^{-\eta b s/v_0}$;
\begin{equation}\begin{split}&
A_H=e^{\frac{\Gamma}{2c}(fx_1 + x_2 -d_2)}\\&
A_V=e^{\frac{\Gamma}{2c}( fx_2+x_1-d_1)}
\end{split}
\end{equation}
As defined in Eq.(\ref{b}), here $b_1=b_2=b$ and $a_1=a_2=0$.
According to Eq.(\ref{dis}), $d_i$ here is
\begin{equation}\label{dis}\begin{split}
d(a=0,b,L_i)  =\left(\frac{c}{\eta b}+L_i\right)\left(e^{\eta
bs/v_0}-1\right)-\frac{cs}{v_0}\end{split}
\end{equation}
Direct calculations show that $|A_V/A_H|-1$ is in the magnitude
order of $10^{-6}$. As shown earlier, the consequence to the final
outcome due to the optical path fluctuation is negligible, therefore
we assume zero fluctuation in the optical paths. Also, since both
photons enter the same Pockels cell, there is no ramping voltage
starting time difference. After calculation, we find that the only
non-constant term in $\Delta\varphi_2 - \Delta\varphi_1$ is
\begin{equation}
\epsilon_2 \approx \frac{k_S\Delta k}{k_{H_1}}x_2
\end{equation}
where $\Delta k = k_{V_2}-k_{H_1}$ which is position dependent. In
the set-up of Ref.\cite{Nature_06_StevensonRM}, the coherence length
of the whole wave train is only about $0.3$m, so the maximal value
of $k_s x_2$ is around 1, also, the magnitude order of $\Delta
k/k_{H_1}$ is  $10^{-3}$, therefore the position-dependent term
$\epsilon_2$ is around $10^{-3}$ and hence negligible.
\subsection{Feasibility}
Similar EOM phase modulation has been used in laser spectroscopy,
such as Pound-Drever-Hall laser frequency
stabilization\cite{black:79}. Technically, such phase modulation can
be accomplished by a commercially available optical device, such as
a Pockels cell, which introduces a phase shift to vertically
polarized mode. In passing through  the Pockels cell, a photon will
acquire an additional phase shift $\alpha V$ given the applied
voltage $V$. Here $\alpha$ is the phase sensitivity of the Pockels
cell. Obviously, $\alpha$ is related to the parameters used in our
earlier calculations by
\begin{equation}
n_0 \eta s =\frac{\alpha \lambda }{2\pi}
\end{equation}
where $\lambda$ is the wave length of the incident light and $n_0$
equals to $c/v_0$. As far as we have known, the phase sensitivity
$\alpha$ of commercially available Pockels cells can be up to $52\
mrad/volt\ @\ 830 nm$\cite{conoptics}. In order to compensate an FSS
of $1\ \mu eV = 2 \pi \times 254.6 \ MHz$, we only need to set $b$
and $b'$ around $30\ V/ns$ according to Eq.~(\ref{vol}). This is
obviously doable by the existing technology. Because the duration of
the field radiation is about several
nanoseconds\cite{PRL_08_StevensonRM}, the scan voltage needs only
last several nanoseconds. Therefore, the maximal voltage requested
is only a few hundred volts according to Eq.~\eqref{vol}, and this
is easily accessible. Moreover,  we can also choose to arrange
several Pockels cells in series along one photon's path to
compensate larger FSS.
 \section{ Concluding remark} We have shown how
to compensate the position dependent phase in the entangled photon
pair generated by the biexciton cascade decay in a
  single semiconductor quantum dot with FSS.  The EOM phase
  modulation is done by  voltage ramping on a Pockels cell.  It is
  shown that our proposed schemes are robust with respect to
  imperfections such as optical path fluctuation. With our scheme,
  the quality of entangled photon pairs can be improved to almost
  perfect level.

\begin{acknowledgments}
{\em Acknowledgments---} We would like to thank H. P. Zeng, C.Z.
Peng, S. Jiang, Jia-Zhong Hu and Ming Gao for helpful discussions. This work
was supported in part by the National Basic Research Program of
China grant nos 2007CB907900 and 2007CB807901, NSFC grant number
60725416, and China Hi-Tech program grant no. 2006AA01Z420.
\end{acknowledgments}


\begin{thebibliography}{23}
  \bibitem{sakurai} J.~J. Sakurai, \textit{Modern Quantum Mechanics},
    (Addison-Wesley Publishing Company, 1994).

  \bibitem{nielsen} M.~A. Nielsen and I.~L. Chuang, \textit{Quantum
    Computation and Quantum Information}, (Cambridge
    University Press, Cambridge, 2000).

  \bibitem{decoy} X.~B. Wang, T. Hiroshima, A. Tomita, and
    M. Hayashi, Physics Reports \textbf{448}, 1~(2007); N.
    Gisin, G. Ribordy, W. Tittel, and H. Zbinden, Rev. Mod.
    Phys. \textbf{74}, 145~(2002); S.~L. Braunstein, and P.
    van Loock, Rev. Mod. Phys. \textbf{77}, 513~(2005).

  \bibitem{PRL_93_KiessTE} T.~E. Kiess, Y.~H. Shih, A.~V.
    Sergienko, and C.~O. Alley, Phys. Rev. Lett.
    \textbf{71}, 3893~(1993).

  \bibitem{PhysRevLett.75.4337} P.~G. Kwiat, K. Mattle, H.
    Weinfurter, A. Zeilinger, A.~V. Sergienko, and Y.~H.
    Shih, Phys. Rev. Lett. \textbf{75}, 4337~(1995).

  \bibitem{Nature_04_EdamatsuK} K. Edamatsu, G.Oohata, R.
    Shimizu, and T. Itoh, Nature \textbf{431}, 167~(2004).

  \bibitem{PRL_04_FattalD} D. Fattal, K. Inoue, J. Vuckovic,
    C. Santori, G.~S. Solomon, and Y. Yamamoto, Phys. Rev.
    Lett. \textbf{92}, 037903~(2004).

  \bibitem{PRL_00_OliverB} O. Benson, C. Santori, M. Pelton,
    and Y. Yamamoto, Phys. Rev. Lett. \textbf{84}, 2513~(2000).

  \bibitem{Nature_06_StevensonRM} R.~M. Stevenson, R.~J.
    Young, P. Atkinson, K. Cooper, D.~A. Ritchie, and A.~J.
    Shields, Nature \textbf{439}, 179~(2006).


  \bibitem{AkopianN_PRL_06} N. Akopian, N.~H. Lindner, E.
    Poem, Y. Berlatzky, J. Avron, D. Gershoni, B.~D.
    Gerardot, and P.~M. Petroff, Phys. Rev. Lett.
    \textbf{96}, 130501~(2006).

  \bibitem{NatPhon_ShieldsAJ_07} A.~J. Shields, Nature
    Photonics \textbf{1}, 215~(2007).

\bibitem{stace2003} T. M. Stace, G. J. Milburn, C. H. W. Baenes, Phys. Rev.
    B \textbf{67}, 085317~(2007).
    \bibitem{guo} Z.~Q. Zhou, C.~F. Li, G. Chen, J.~S. Tang, Y.
  Zou, M. Gong, and G.~C. Guo, arXiv:0909.0078v1~(2009).

  \bibitem{PRL_07_HudsonAJ} A.~J. Hudson, R.~M. Stevenson,
    A.~J. Bennett, R.~J. Young, C.~A. Nicoll, P. Atkinson,
    K. Cooper, D.~A. Ritchie, and A.~J. Shields, Phys. Rev.
    Lett. \textbf{99}, 266802~(2007).

  \bibitem{PRL_08_StevensonRM} R.~M. Stevenson, A.~J.
    Hudson, A.~J. Bennett, R.~J. Young, C.~A. Nicoll, D.~A.
    Ritchie, and A.~J. Shields, Phys. Rev. Lett.
    \textbf{101}, 170501~(2008).

  \bibitem{NJP_06_YoungRJ} R.~J. Young, R.~M. Stevenson, P.
  Atkinson, K. Cooper, D.~A. Ritchie, and A.~J. Shields, New
  Journal of Physics \textbf{8}, 29~(2006).

\bibitem{NJP_07_HafenbrakR} R. Hafenbrak, S.~M. Ulrich, P.
  Michler, L. Wang, A. Rastelli, and O.~G. Schmidt, New
  Journal of Physics \textbf{9}, 315~(2007).

\bibitem{he:157405} L. He, M. Gong, C.-F. Li, G.-C. Guo, and
  A. Zunger, Phys. Rev. Lett. \textbf{101}, 157405~(2008).

\bibitem{PRL_09_YoungRJ} R.~J. Young, R.~M. Stevenson, A.~J.
  Hudson, C.~A. Nicoll, D.~A. Ritchie, and A.~J. Shields,
  Phys. Rev. Lett. \textbf{102}, 030406~(2009).
  \bibitem{jones} Nick S. Jones and T. M. Stace, Physical Review A \textbf{73},
033813 (2006)
\bibitem{np}S. Barrett, Nature Photonics, {\textbf 3}, 430 (2009)
\bibitem{PhysRevLett.76.3005} D. Gammon, E.~S. Snow, B.~V.
  Shanabrook, D.~S. Katzer, and D. Park, Phys. Rev. Lett.
  \textbf{76}, 3005~(1996).

\bibitem{PhysRevB.65.195315} M. Bayer, G. Ortner, O. Stern,
  A. Kuther, A.~A. Gorbunov, A. Forchel, P. Hawrylak, S.
  Fafard, K. Hinzer, T.~L. Reinecke, S. N. Walck, J. P.
  Reithmaier, F. Klopf, and F. Schafer, Phys. Rev. B
  \textbf{65}, 195315~(2002).

\bibitem{scully} M.~O. Scully and M.~S. Zubairy,
  \textit{Quantum optics} (Cambridge University Press, Cambridge, 1997).

\bibitem{black:79} E.~D. Black, American Journal of Physics
  \textbf{69}, 79~(2001).

\bibitem{conoptics} Http://www.conoptics.com/Modulation-Systems-Products.html.





\end{thebibliography}

\end{document}